# Randomness of imperfectly entangled states.

Myriam Nonaka, Mónica Agüero, Marcelo Kovalsky and Alejandro Hnilo
*CEILAP, Centro de Investigaciones en Láseres y Aplicaciones, UNIDEF (MINDEF-CONICET);*
*CITEDEF, J.B. de La Salle 4397, (1603) Villa Martelli, Argentina.*
email: ahnilo@citedef.gob.ar
August 28th, 2019.

The generation of series of random numbers is an important and difficult problem. Appropriate measurements on entangled states have been proposed as the definitive solution, based on the impossibility of exploiting quantum non-locality to get faster than light signaling. There is a controversy regarding what is preferable to produce series with utilizable randomness in practice: high or low entanglement. We prepare biphotons with three different levels of entanglement: "easy" entangled ($S_{CHSH}$ =2.67), marginally entangled ($S_{CHSH}$ =2.06), and no-entangled ($S_{CHSH}$ =1.42). Randomness is evaluated, independently of the quantum non-locality argument, through a battery of standard statistical tests, Hurst exponent, Kolmogorov complexity, Takens' dimension of embedding, and Augmented Dickey-Fuller and Kwiatkowski–Phillips–Schmidt–Shin tests to check stationarity. The no-entangled case is found to produce the smallest rate of not-random series, and the marginal case the largest. Although the entangled case has a larger rate of not-random series than the no-entangled case, it is found still acceptable for QKD.

## 1. Introduction.

Series of random numbers are a basic supply in many applied sciences of information. The generation and evaluation of randomness is difficult. Even the very definition of "random" is difficult. In a rough classification, *Borel normality* means that the frequency of strings of 1 and 0 of different length (in a binary sequence) is statistically equivalent to the one that would be expected by tossing an ideal coin. Other ways to evaluate randomness measure the decay of self-correlation or mutual information. They all involve statistics and require the property of stationarity. This means that the probability distributions, or at least the covariances, are constant along the series. The battery of tests provided by the National Institute of Standards and Technology (NIST) [1] is mostly based on these approaches. *Algorithmic complexity* means that there is no program code able to generate the series by using a number of bits shorter than the said series. Note that this definition does not use probabilities. It applies even to sequences that are not stationary. An algorithmically complex sequence is demonstrated to be *non-computable* and Borel normal, but the inverses are not true [2]. Yet, algorithmic (or Kolmogorov) complexity cannot be truly calculated. It can only be estimated.

Some consensus has been reached, that appropriate measurements performed on quantum systems "certify" randomness. This consensus has two bases: one is von Neumann's axiom, which states that quantum measurements violate Leibniz's principle of sufficient reason. In other words: that a quantum measurement produces one or another outcome *without cause*. A sequence of such outcomes is intuitively random, but this intuition is difficult to formalize [3]. The other basis is that, assuming the existence of quantum non-locality (in observations performed on spatially extended entangled states), the sequence must be unpredictable to prevent faster than light signaling [4]. Following this argument, a series produced by measurements made on a spatially extended entangled state is non-computable [5,6]. Nevertheless, these bases are not free of doubts: von Neumann's axiom may be understood as a description of how to use and interpret Quantum Mechanics (QM), of what QM can or cannot do, but not of properties of Nature. Also, non-locality is alien to the QM usual (Copenhagen) interpretation. This interpretation denies existence of a physical reality independent of the observer, but it does *not* consider the existence of non-local effects.

Randomness of the series produced by measurements on spatially extended entangled states is crucial for the security of Quantum Key Distribution (QKD). Following the "quantum certification" argument, the purity of the achieved entanglement puts a minimum bound on the entropy of the generated series [7] and hence, to the degree of statistical randomness. Loophole-free verification of the violation of Bell's inequalities has been required as a necessary step to certify randomness [3]. Loophole-free setups have been recently used to produce series with quantum certified statistical randomness [8,9]. Yet, algorithmic randomness of quantum-produced series remains controversial. An experimental approach has been proposed to explore this problem [3,10].

The necessity of an experimental exploration, even if the idea of quantum certified randomness is entirely accepted, is strengthened by the observation that series produced by quantum devices often show a poor level of randomness in practice [11-13]. These failures are supposed to be caused by technical imperfections. Series generated in highly refined (extremely difficult to perform), loophole-free setups do pass all known tests of randomness [14], but they are too short, and too cumbersome to produce, to be of practical use. It is then natural asking to what extent quantum-based random numbers generators are practical and reliable at the current technological level. In order to answer this question an evaluation of randomness of the produced series, independent of the quantum non-locality based argument, is in order.

In order to generate an entangled state, the non-distinguishability of the two paths in an interference phenomenon must be achieved, and stabilized. This is a main technical difficulty. Knowing what happens under conditions of imperfect non-distinguishability is hence of direct practical interest. In this paper, we study the influence of distinguishability on randomness. It is worth mentioning here the theoretical study on the effect of nonlocality on randomness [15], which concludes that QM certifies maximal randomness even if non-locality and entanglement are not maximal (what is somehow contrary to intuition). Also, the recent experimental result using high purity states of different Concurrence [16] showing that, in practice, the presence of noise implies that lower entanglement leads to lower randomness, in agreement with intuition. Our contribution here deals with mixed states instead, so it presents complementary results.

By adjusting the distinguishability of the state's components, we prepare states of different levels of entanglement and Purity and evaluate the randomness of the produced series. The setup uses pulsed pumping, which is close to the intuitive method of "tossing a coin" to produce random numbers, and time stamped record of detections, which gives freedom to compose different types of series with the same experimental data. It also allows stroboscopic reconstruction of time variation of magnitudes of interest. It is described in the Section 2.

Randomness of a series cannot be demonstrated. What can be demonstrated is that a series is *not-random*. Hence, in order to evaluate randomness independently of the quantum non-locality argument, we count the number of series (in a set) rejected by usual tests. We apply the full battery of tests of the NIST, calculate Hurst exponent, look for a compact object in phase space (Takens' reconstruction theorem) and use standard Augmented Dickey-Fuller (ADF) and Kwiatkowski–Phillips–Schmidt–Shin (KPSS) tests to check stationarity. For the Reader not familiar with these indicators, they are briefly described in the Appendix.

Obtained results are summarized in Section 3 Complete time series mean several Gb of data, which we are glad to share upon request. We measure the rate of not-random series produced in no-entangled, marginally entangled, and entangled cases. Intuitively, the first case is expected to be "more random" because of the larger presence of noise (in other words: a heavier weight of the unpredictable environment). The latter is expected to be "more random" for it is closer to the condition of quantum certified randomness.

Our results, in few words: the no-entangled case has the lowest rate of not-random series, while the intermediate case has the highest. This seems to be good news, because no-entangled states are easier to get and reproduce than entangled ones. But, of course, entangled states are necessary for QKD. In this sense, an important result is that the entangled case, although not the "most random" one in the studied set, has a rate of not-random series which is, *prima facie*, still acceptable for QKD.

## 2. Experimental setup.

Biphotons at 810 nm entangled in polarization in the fully symmetrical Bell state $|\phi^+\rangle$ are produced in the standard configuration using two crossed 1mm length each BBO-I crystals [17], pumped by a 40 mW diode laser at 405 nm, see Figure 1. This laser can be modulated at will (bandwidth 20 MHz). Unless stated otherwise, files are obtained with square shaped pulses 500 ns long at a 50 Khz rate, (2.5% duty cycle), the run lasts 300s for each angle setting $\{a,b\}$ of the polarizers. A time tagged file produced in these conditions has $\approx 1.2 \times 10^6$ single counts in each station.

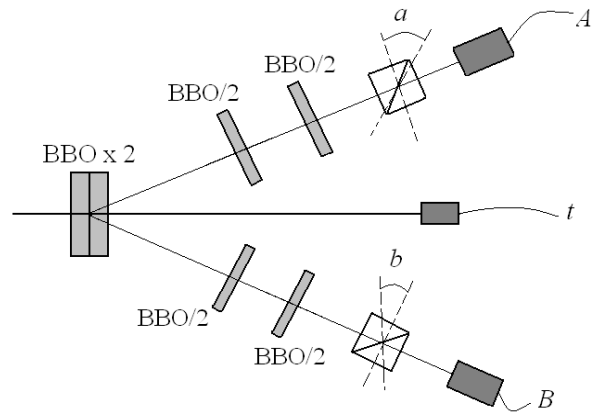

Figure 1: Sketch of the setup. A pulsed laser beam at 405 nm pumps a set of crossed BBO crystals producing two beams of radiation at 810 nm entangled in polarization. Half-length crystals are inserted in the path of these beams to compensate walk-off. By inserting or removing these crystals, states of different levels of entanglement and purity are prepared. Polarizers are oriented at angles *a* and *b*. Single photons are detected at *A* and *B*. The start of the pump pulse is registered with photodiode *t*. Signals $\{A,B,t\}$ are recorded in a time-to-digital converter with 1 ps nominal resolution.

Longitudinal walk-off is not an issue, because of the laser's coherence length, nominally longer than 20m (measured > 4 cm). Transverse walk-off is compensated with two additional BBO crystals of half length, placed on each branch. By inserting or removing these crystals the degree of distinguishability and hence the level of entanglement and Purity of the final state is adjusted. We discuss here three cases, identified by the measured value of the Clauser-Horne-Shimony and Holt (CHSH) parameter S: entangled S=2.67, marginally entangled S=2.06, no-entangled S=1.42. The first is chosen to represent a state easy to reproduce in a rough device of practical use. The latter is chosen to fit the value (S=√2) of the semiclassical radiation theory. Concurrence (*C*) and Purity (*P*) are also measured to characterize the state.

Photons are detected in each station with avalanche photodiodes. The time values of detection of each photon are stored in a time-to-digital converter with nominal resolution 1 ps, limited by detectors' jitter to ≈1 ns. Two time series are recorded for each angle setting. Two sets of settings $\{a=0, a'=2\theta, b=\theta, b'=3\theta\}$ are used in each case: $\theta=\pi/8$ for the CHSH inequality, and θ=8.6° for the algorithmic [18] and informatic [19] ones.

In addition, pulsed pump and time-stamped record of data allow the stroboscopic reconstruction of the time evolution of parameters of interest. As an illustration, Figure 2 shows the time evolution of the singles and coincidences for the setting $a=0$, $b=\pi/8$. Other 15 files recorded with the appropriate settings allow the stroboscopic reconstruction of the time evolution of the parameter S and averaged efficiency. There are ≈0.06 photons detected per pulse in the average. The number of coincidences depends on the angle setting; the average efficiency is ≈0.25.

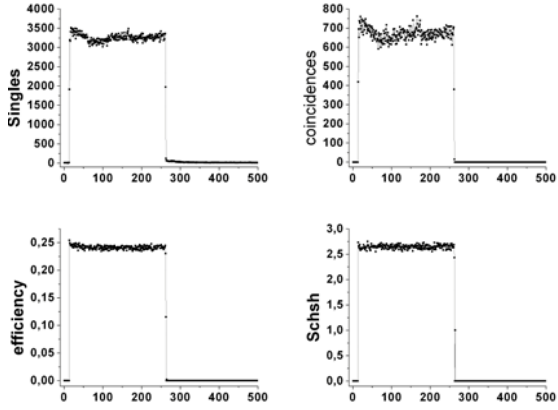

Figure 2: Stroboscopic time evolution (2ns per unit) of: singles, station A (setting: $a=0$), coincidences (setting: $a=0$, $b=\pi/8$), efficiency averaged over all settings (station A), S parameter. Singles and coincidences show a deviation from perfect square shape which is observed with a fast photodiode too. Curves of efficiency and S involve ratios of singles and coincidences and are free of that deviation.

Figure 2 has interest because, in a previous experiment performed with a different laser and poorer time resolution, a linear increase of the efficiency with time was observed [20]. This effect might have had both practical and fundamental consequences. The effect is not observed now; efficiency is constant during the pulse. Note also the time stability of the S parameter. Even small deviations from perfect square shape (which do exist in the pump pulse) disappear.

## 3. Observed randomness.
*3.1 Types of generated time series.*

The setup in Fig.1 generates a huge amount of raw data, which can be arranged in different ways. In this paper, we use only times of *coincidences* between the two stations to compose the following types of series:

#1) Real time between successive coincidences: files named dt* (* indicates the setting).

#2) Time distance of the coincidence to the start of the pump pulse, defined by the trigger signal from the photodiode: files named deltat*.

Figure 3 explains the meaning of these two types of series. Two sources of randomness are hence combined in type #1: the time elapsed within the pump pulse where the coincidence occurs, and the time elapsed between successive pulses where a coincidence occurs, which is affected by pulse jitter. In type #2 instead, only the former source of randomness is involved.

It is worth recalling that series of time detections allow reconstructing series of outcomes, at least in QKD setups [21].

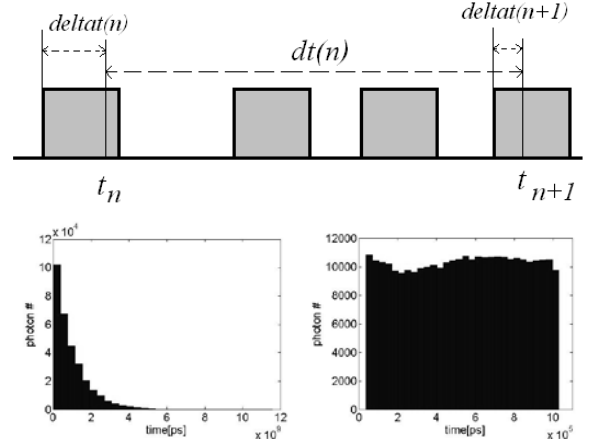

Figure 3: About series types #1 and #2. Up: Pump pulses are indicated with grey rectangles, duration and separation are not in scale and jitter is exaggerated. Consider two coincidences, detected at times $t_n$ and $t_{n+1}$. Time elapsed between successive coincidences (*dt*) builds type #1 series. Time elapsed from the start of each pump pulse (*deltat*) builds type #2 series. Down: typical histograms (number of coincidences per time slot) of type #1 (left) and #2 (right). Type #1 has the exponential shape corresponding to a Poisson distribution [11]. Type #2 fills the interval of the pulse duration copying the pump pulse shape (compare with Figs.2a,b). Histograms correspond to settings $a=0$, $b=22.5°$, pulse duration 1μs, rate 50 Khz, run lasted 300 s, S=2.24.

Series types #1 and #2 are directly analyzed with Takens' reconstruction method and Hurst exponent. In order to generate binary series suitable to be analyzed with the other tests, a "1" ("0") corresponds to a time difference above (below) some threshold value. The obtained series are different depending on this value. Following [11], we use both the time average and the median of the distribution as threshold values, and check the difference. The resulting binary series are then analyzed with the NIST battery, Kolmogorov complexity, ADF and KPSS.

#3) Outcomes: By intercalation of dt* files of settings with angles at 90°, it is possible to get binary series of outcomes as if there were two detectors per station. If a coincidence is recorded in a rotated (not-rotated) setting, a "1" ("0") is written in the series of outcomes. As a consequence, the number of series halves and their length doubles. This procedure has been used even in setups which actually have two detectors per station, to get balanced series despite the problem of different detectors' efficiencies [18]. Type #3 series are then naturally free of this problem.

The intercalation method is based on a "strong counterfactual" assumption: observed and unobserved outcomes are assumed to be the same and to occur at the same time values if their nature (observed or unobserved) is flipped. The equivalent assumption in usual Bell's inequalities is weaker: that only the *averages* are equal [22]. The "strong counterfactual" has been used to derive the Clauser-Horne inequality from arithmetical properties of the series only, without any reference to Locality or Realism [23,24]. As discussed below, the intercalation method apparently introduced deviation from randomness.

The main results are discussed in the following subsections.

*3.2 Results for the entangled case (S=2.67, C=0.86, P=0.87).*

A total of 72 series are analyzed, 32 with $\theta=22.5^o$ and 40 with $\theta=8.6^o$.

The ADF test indicates that unit-root can be discarded in all series, while the KPSS test rejects stationarity in two series (both type #2, one with $\theta=22.5^o$ and the other with $\theta=8.6^o$, they do not correspond to the same setting).

We consider that a series is rejected by the NIST battery if it does not pass just one of the *applicable* tests (some of them are not reliable if the series is too short). Two series are rejected in this way: one with $\theta=8.6^o$ type #1 and one with $\theta=22.5^o$ type #2, *and all the type #3 series*.

| File type | $Kc$ | $Km$ | H |
|---|---|---|---|
| #1, $\theta=22.5^o$ | 0.947 ± 0.004 | 1.020 ± 0.004 | 0.497 ± 0.016 |
| #2, $\theta=22.5^o$ | 1.022 ± 0.010 | 1.019 ± 0.004 | 0.493 ± 0.013 |
| #1, $\theta=8.6^o$ | 0.966 ± 0.007 | 1.022 ± 0.008 | 0.493 ± 0.020 |
| #2, $\theta=8.6^o$ | 1.020 ± 0.007 | 1.021 ± 0.007 | 0.491 ± 0.016 |
| #3, $\theta=8.6^o$ | 1.015 ± 0.001 | not applicable | 0.505 ± 0.009 |

Table 1: Average values (over all files of the same type) and dispersion of Kolmogorov complexities and Hurst exponent, for files recorded with entangled state, S=2.67.

As it will be seen, the rejection of all type #3 series by the NIST battery occurs in the three reported cases (entangled, marginally entangled, no-entangled). The cause must be in the intercalation method, but it is not evident why. Each type #3 series is rejected by more than one test, and the "rejecting tests" are not the same for all series. We consider the rejection of type #3 series by the NIST battery as a special case that deserves further study.

Averaged (over the set of files) and dispersion values of Kolmogorov complexities and Hurst exponent are summarized in Table 1. If the binary series is composed using the average value (the median) the complexity value is $Kc$ ($Km$) [11]. Hurst exponent is calculated directly for the time series. We use the criterion that the set of files is complex and uniformly random if the ideal value (in each case: 1 and ½) is within the dispersion range. According to this criterion, all sets of files in Table 1 are random.

Takens' reconstruction method finds a compact object in phase space with $d_E$ ranging between 6 and 8, in 4 series (3 type #1, 2 of them with $\theta=22.5^o$, the remaining one is type #2 with $\theta=22.5^o$). All of them have at least one positive Lyapunov exponent. It means strong sensitivity to variations, which is one of the signatures of chaos. This explains why these series are "apparently" random and are able to pass the other tests. This method cannot be applied to type #3 series.

In summary, leaving aside the 8 type #3 series (which are all rejected by NIST and are considered a special case), 8 out of the remaining 64 are found not-random by one of the used criteria. No series is rejected by more than one criterion. Therefore, all the tests act in a complementary way, as planned.

*3.3 Results for the marginally entangled case (S=2.06, C=0.62, P=0.67).*

A total of 72 series are analyzed, 32 with $\theta=22.5^o$ and 40 with $\theta=8.6^o$.

ADF indicates that unit-root can be discarded in all series, while KPSS discards stationarity in 5 (2 type #1 and 2 type #2 all with $\theta=22.5^o$, and one type #1 with $\theta=8.6^o$, settings are all different). NIST rejects 5 series (none is the same as before, one is type #2 with $\theta=22.5^o$, and 4 are type #1 with $\theta=8.6^o$, settings are all different). As in the previous subsection, NIST also rejects all the 8 type #3 series.

Averaged (over the set of files) and dispersion values of Kolmogorov complexities and Hurst exponent are summarized in Table 2. According to the criterion stated in the previous subsection, all the sets of files in Table 2 are random.

| File type | $Kc$ | $Km$ | H |
|---|---|---|---|
| #1, $\theta=22.5^o$ | 0.965 ± 0.005 | 1.031 ± 0.025 | 0.501 ± 0.011 |
| #2, $\theta=22.5^o$ | 1.019 ± 0.004 | 1.023 ± 0.015 | 0.498 ± 0.009 |
| #1, $\theta=8.6^o$ | 0.965 ± 0.008 | 1.024 ± 0.013 | 0.504 ± 0.021 |
| #2, $\theta=8.6^o$ | 1.020 ± 0.006 | 1.021 ± 0.007 | 0.499 ± 0.010 |
| #3, $\theta=8.6^o$ | 1.019 ± 0.010 | not applicable | 0.503 ± 0.011 |

Table 2: Average values and dispersion of Kolmogorov complexities and Hurst exponent, for files recorded with marginally entangled state, S=2.06.

Takens' reconstruction method finds a compact object in phase space with $d_E$ ranging between 6 and 9, in 4 series (3 type #1, one of them with $\theta=22.5^o$, the other two with $8.6^o$, the remaining one is type #2 with $\theta=8.6^o$, settings are all different). In all of them, at least one Lyapunov exponent is positive, as before.

In summary, leaving aside the 8 type #3 series (rejected by NIST), 14 out of the remaining 64 are found not random by one of the used criteria. None is discarded by more than one criterion.

*3.4 Results for the non-entangled case (S=1.42, C=0.44, P=0.56).*

A total of 72 series are analyzed in this case, 32 with $\theta=22.5^o$ and 40 with $\theta=8.6^o$.

As before, ADF indicates that unit-root can be discarded in all series. Instead, KPSS cannot discard stationarity in any series. NIST rejects no series excepting the 8 type #3, as in the two previous cases.

Curiously, to this case belongs the only series (a type #3 one), among the 216 studied, with relatively low complexity, $Kc$ =0.769. This is the reason of the higher dispersion in this set, see Table 3. This is also the only series discarded by more than one of the used tests.

| File type | $Kc$ | $Km$ | H |
|---|---|---|---|
| #1, $\theta$=22.5° | 0.962 ± 0.002 | 1.016 ± 0.003 | 0.496 ± 0.013 |
| #2, $\theta$=22.5° | 1.016 ± 0.003 | 1.017 ± 0.004 | 0.497 ± 0.012 |
| #1, $\theta$=8.6° | 0.964 ± 0.009 | 1.018 ± 0.010 | 0.496 ± 0.022 |
| #2, $\theta$=8.6° | 1.017 ± 0.007 | 1.016 ± 0.008 | 0.496 ± 0.012 |
| #3, $\theta$=8.6° | 0.972 ± 0.086 | *not applicable* | 0.501 ± 0.008 |

Table 3: Average values and dispersion of Kolmogorov complexities and Hurst exponent, for files recorded with no entangled state, S=1.42.

Takens' method is not able to find a compact object in phase space in any series. This is probably due to the high level of noise intrinsic to the state's lower correlation. Noise means coupling with the infinite dimensions of the environment and, naturally, to a non-measurable value of $d_E$. In consequence, Lyapunov exponents cannot be calculated in this case.

In summary, leaving aside the 8 type #3 series (all rejected by NIST), 0 out of the remaining 64 are found not random by one of the used criteria.

**Comments.**

The main purpose of this paper is to experimentally study the impact of no-distinguishability (which is a practical difficulty in the preparation and stability of an entangled state) in the randomness of series generated detecting biphotons. Randomness is not derived from a non-locality argument, but independently "measured" by the rate of rejected (or not-random) series, using several well established methods. Leaving aside the 24 type #3 series, which are all discarded by the NIST battery test for reasons still to be elucidated, a tight summary of our results is that 8 out of 64 series recorded for S=2.67 are not random, 14 out of 64 for S=2.12, and 0 out of 64 for S=1.42. The number of series we analyze is still scarce, in statistical terms, to extract definitive conclusions, but it suffices to draw a tendency to guide future research.

A pertinent question is whether the level of not-randomness of the entangled state is tolerable for QKD. A simple criterion is to check whether this level is higher than the probability (of guessing future settings) a classical model needs to reproduce QM predictions. Depending on the type of classical model assumed, this value is between 0.14 [25] and 0.25 [26] above the minimum of ½ for the usual setup with two possible settings. As the level of not-randomness is below (8/64 = 0.125 < 0.14) that probability, a classical model is not able to reproduce QM results in this case. Hence, the (easily reproducible in practice) studied entangled state has a level of not-randomness that is still acceptable for QKD (strictly speaking, the numbers mean that it cannot be discarded by this argument). The conclusion changes if the total rate of rejected series (16/72) is taken into account instead. But, the poor result obtained with type #3 series is probably caused by some fault in the intercalation method (a fault to which the NIST battery is especially sensitive). The origin of the fault is not evident and deserves further research.

The marginally entangled state produces the largest ratio of rejected series. This is mostly due to the relatively large number of non-stationary (trend-stationary) series. It is remarkable that no series is found to be unit-root. Also, that no series obtained with S=1.42 reveals a compact object in phase space, what is probably caused by the high level of uncorrelated coincidences ("noise", high-dimension dynamics) in this case.

Taking into account the obtained results and the facility of its realization, a no-entangled, noisy source is the best candidate for a practical random number generator. But, of course, it is not applicable to QKD, where entanglement is unavoidable. According to our results, an easily reproducible entangled state provides a level of not-randomness which is not sufficient for a classical model to reproduce the observations. As these observations are hence classically unpredictable, we consider this state, in spite of not being "fully random", still useful for QKD.

**Appendix.**
*Indicators of randomness (NIST, Hurst, complexity, Takens' $d_E$, ADF and KPSS).*

The battery developed by the NIST consists of 15 different tests. Not all of them can be applied in all cases, because of series' length. The battery and its details are available in the NIST's page [1], so it is unnecessary repeating their description here. We just mention that it essentially checks Borel normality, hidden periodicities, and decay of mutual information.

The Hurst exponent is related to the rate at which autocorrelations decay. It is usually named H, and normalized between 0 and 1. H > ½ means the series has long range correlations, H < ½ that it has strong fluctuations in the short term, while H ≈ ½ means that it is uniformly random.

Complexity (Kolmogorov) has advantages over other methods of detecting regular behavior. It does not need, at least in principle, to assume stationary probabilities (see below). It applies to series of any length. On the other hand, complexity cannot be actually *computed*, for one can never be sure that there is no shorter program (than the one that is already found) able to generate the series. Complexity can only be *estimated* from the rate of compressibility of the series using, f.ex., the algorithm devised by Lempel and Ziv [27]. Here we use the approach developed by Kaspar and Schuster [28] and implemented by Mihailovic [29] to estimate normalized complexity K. This value is designed to be near to 0 for a periodic or regular sequence, and near to 1 for a random one. For relatively short and strongly fluctuating series, values K>1 may occur. Note that this procedure is ultimately statistical, too.

Nonlinear analysis of series [30] provides a completely different approach. In a chaotic system, for example, a few dynamical variables are linked through nonlinear equations in such a way that the evolution is apparently random. Nevertheless, the evolution involves few degrees of freedom and is partially predictable (there is a finite *horizon of predictability*, roughly given by the inverse of the largest positive Lyapunov exponent). This is a fundamental difference with "true" random evolution, which can be thought of as requiring a very high (eventually, infinite) number of degrees of freedom to be described. Takens' reconstruction theorem and related methods allow measuring the number of dimensions of the object in phase space within which the system evolves, and hence to discriminate chaos from randomness. That number is called *dimension of embedding*, $d_E$. A definite value of $d_E$ (which is always small compared with the series' length) indicates the series is not random. In some cases this approach allows the prediction of future elements of the series (within the horizon of predictability), which is the ultimate proof of not-randomness. This method was able to reveal the existence of regularities in one of the runs of the Innsbruck experiment [21,31].

Excepting for Takens' $d_E$, all the mentioned methods are statistical, and hence require the series to be stationary. It is then crucial to ensure stationarity to apply any of the statistically based methods. There are two main types of *non*-stationarity. One: the series' statistical parameters follow a continuous and slow evolution (*trend-stationary*). Deviations from the average trend vanish as the number of elements in the series increases. By identifying and correcting the trend, the series can be made stationary again. Two: a deviation affects the values of the statistical parameters in a permanent way through the series (*unit-root*). Standard tests for the two types exist (KPSS and ADF) but, because of the very nature of the involved hypotheses and methods, they do not provide definitive conclusions. Used together, they *indicate* the most probable nature of the series. KPSS tests the null hypothesis that the series is trend-stationary, against the alternative of unit-root. Obtaining "0" (1) indicates that stationarity cannot (can) be rejected. ADF tests unit-root. Obtaining "0" (1) indicates that unit-root cannot (can) be rejected.

**Acknowledgments.**

Many thanks to Prof. Dragutin Mihailovic for his help to use the algorithms to estimate Kolmogorov complexity and to interpret their outputs. This work received support from the grants N62909-18-1-2021 Office of Naval Research Global (USA), and PIP 2017 0100027C CONICET (Argentina).